\documentclass[a4paper]{jpconf}
\usepackage{graphicx}
\usepackage{epsfig}
\usepackage[T1]{fontenc}
\usepackage{amssymb}
\usepackage{amsmath}
\usepackage{amsfonts}
\usepackage{verbatim}
\usepackage{graphicx}
\usepackage{hyperref}
\usepackage{array}
\usepackage{multirow}
\usepackage{epstopdf}
\usepackage{color}  

\begin{document}
\title{High speed fault tolerant secure communication for muon
chamber using fpga based gbt emulator}

\author[1]{Suman Sau$^{1}$, Swagata Mandal$^{2}$, Jogender Saini$^{2}$, Amlan Chakrabarti$^{1}$ and Subhasis Chattopadhyay$^{2}$}
\address{$^{1}$A.K Choudhury School of IT, University of Calcutta $^{2}$Variable Energy Cyclotron Centre, Kolkata, India}

\ead{$^{1}$ssakc\_s@caluniv.ac.in, $^{2}$swagata.mandal@vecc.gov.in}
\vspace{-1 pc}
\begin{abstract}
The Compressed Baryonic Matter (CBM) experiment is a part of the Facility for Antiproton and Ion Research (FAIR) in Darmstadt at the GSI. The CBM experiment will investigate the highly compressed nuclear matter using nucleus-nucleus collisions. This experiment will examine heavy-ion collisions in fixed target geometry and will be able to measure hadrons, electrons and muons. CBM requires precise time synchronization, compact hardware, radiation tolerance, self-triggered front-end electronics, efficient data aggregation schemes and capability to handle high data rate (up to several TB/s). 
 As a part of the implementation of read out chain of MUCH in India, we have tried to implement FPGA based emulator of GBTx in India. GBTx is a radiation tolerant ASIC that can be used to implement multipurpose high speed bidirectional optical links for high-energy physics (HEP) experiments and is developed by CERN. 
GBTx will be used in highly irradiated area and more prone to be  affected by multi bit error. To mitigate this effect instead of single bit error correcting RS code we have used two bit error correcting (15, 7) BCH code. It will increase the redundancy which in turn increases the reliability of the coded data. So the coded data will be less prone to be affected by noise due to radiation. Data will go from detector to PC through multiple nodes through the communication channel. In order to make the data communication secure, advanced encryption standard (AES - a symmetric key cryptography) and RSA (asymmetric key cryptography) are used after the channel coding. We have implemented GBTx emulator on two Xilinx Kintex-7 boards (KC705). One will act as transmitter and other will act as receiver and they are connected through optical fiber through small form-factor pluggable (SFP) port. We have tested the setup runtime using Xilinx Chipscope Pro Analyzer. Also measure the resource utilization, throughput, power optimization of implemented design. 
\end{abstract}
\vspace{-3 pc}
\section{Introduction}
\label{Introduction}
 High speed and fault resilient DAQ system is an integral part of the signal processing unit in real time applications like HEP, satellite communication etc. Traditional DAQ system front end electronics (FEE) board captures data from the sensors through high speed LVDS link, process it and sends it to storage device using high speed link like Ethernet, PCIe, fiber optic etc. In~\cite{Wang:iciea:2009}, a slow data rate (1.6 Gbps) DAQ architecture is described. In~\cite{kadric:socc:2012}, authors show a high speed (8.5Gbit/s) optical communication between two ALTERA Stratix IV FPGA boards. Here, PCIe buses are used to transfer data between board and computing node, but no error correction or cryptographic mechanism is used. 
In Beijing Spectrometer III (BESIII) trigger system, a high speed data transmission protocol over optical fiber for real time data acquisition was developed by Hao Xu \textit{et.al} in~\cite{haoxu:nss:2007:XX}. The system used Multi Gigabit Transceiver (MGT) of Virtex-II Pro series FPGA for optical fiber communication (1.75 Gbps). In ~\cite {mattihalli:cecnet:2012},~\cite{bohm:nss:mic:2012} a high speed data transfer protocol was implemented using different FPGAs, which achieved highest data rate of 2.5 Gbaud and 784 Mbps receptively.  
\par Single event upset (SEU) occurs when a charged particle hits and transfers sufficient energy to the silicon area of a circuit. Two type of approaches are taken for the SEU mitigation: prevention and recovery. Prevention methods are mainly considered during the ASIC design. In recovery methods different online recovery mechanisms \textit{e.g.} fault tolerant computing, error detecting/correcting code and online testing are used. The concurrent error detection (CED)~\cite{Siewiorek:CED} is one of such techniques where an extra error detection circuit is attached with the main circuit that simply recomputes or rolls back the whole operation from the beginning when an error is detected. In Triple Modular Redundancy (TMR)~\cite{TMR:IeeeTran:NuclrPhy} the same functional replica is used thrice and final result is taken with the majority voting system.

\par
In the above mentioned papers the author did not propose any idea to handle the SEU mitigation in high speed data acquisition system that are used in an adverse environmental condition as found in HEP experiment. For data communication among different nodes secure transmission is also an issue. In real MUCH experiments environment will be as follows
\begin{itemize}
\item \# Channels >100k
\item Read out Frequency > 100 KHz
\item Synchronization limit <100 ps
\item Data Capacity >1 Tb/s
\end{itemize}
In our paper, we proposed a high speed data acquisition system design with secure communication using Gigabit Transmitter(GBT)~\cite{GBT:ref1}emulator, which is protected from SEU by multi-bit error correcting BCH code~\cite{chien:search}  and interleaver. Scrambling is used as line coding technique to maintain the DC balance and to obtain 20\% extra throughput a compared to 8b/10b coding in~\cite{Minami:ieeetran:2011},~\cite{kadric:socc:2012},~\cite{Liu:i2mtc:2013:X}. We have achieved maximum data rate of 4.8 Gbps compared to 1.6 Gbps, 2.125 Gbps, 1.75 Gbps in~\cite{Minami:ieeetran:2011},~\cite{Liu:i2mtc:2013:X},~\cite{haoxu:nss:2007:XX} respectively.
In this paper, our key contributions are:
\begin{itemize}
\item Efficient implementation of high speed
secure DAQ with optical link for muon experiment using FPGA based GBT emulator.
\item DAQ system is protected from single event upset (SEU)
by multi-bit error correction technique
\item Implementing cryptographic algorithms in hardware level to ensure secure PC communication
\item Interfacing the DAQ with PC through PCIe (Generation
2, Lane 8) and scatter gather direct memory access
(SGDMA)

\end{itemize}   
The rest of the paper is organized as follows. Section~\ref{SystemDesignDAQ} describes the full system design topology for the high speed secure DAQ. Experimental setup with performance evaluation are described in Section~\ref{PerformenceEvoluation} followed by concluding remarks in Section~\ref{Conclusion}.
\vspace{-1 pc}

\section{High speed DAQ design with secure communication for MUCH experiments}\label{SystemDesignDAQ}
High data rate, error correction capabilities, secure communication and efficient storage mechanism remains the main criteria of DAQ design for MUCH experiments for future analysis. In our system for high data transmission, optical fiber is used as the communication media. Several multibit error correction methods for efficient communication had been discussed in Section~\ref{Introduction}, where BCH coding is most suitable for random error correction. The interleaver block has been introduced after encoder block judiciously to enhance the error correction efficiency. For ensuring secure communication, different cryptographic algorithms can be used. We have chosen RSA and AES algorithms which are discussed in section~\ref{security} for test our prototype. In the receiver side data is directly transfered to PC through PCIe from the FPGA board. Functional blocks of the proposed system is shown in Figure~\ref{fig:BlockDiagramFlow}. The details of each block have been discussed in the following subsections. 

\vspace{-3 pc}
\subsection{Scrambler/Descrambler} Occurrence of long sequences of `1' (or `0') are reduced by the scrambler. This process helps to maintain the DC balance in input signal coming from the detector/sensor and also helps in accurate timing recovery on receiver side. Scrambler does not add any overhead in the system like the $8b/10b$ or $7b/8b$ line coding except a latency of one clock cycle. In our system, 52 bit incoming data is divided into four blocks of 13 bit data and each block scrambled simultaneously using 13 bit polynomial. 
\vspace{-1 pc}
\subsection{BCH Encoder/Decoder}A binary error correcting code BCH (15,7,2) code is used to correct the error due to SEU or MBU. Here (15,7,2) means, 7 bit of input data, 8 redundant bit which are appended with the input data and it can correct up to two bit of error. So the code rate (ratio of input data to coded data) is 0.467. In the proposed DAQ, 56 bit of data (52 bit data with 4 bit header) are broken down into eight 7 bit of data, which are encoded with BCH encoder in parallel. After encoding the 15 bit of data from the eight blocks, they are assembled to generate a total 120 bit data. This block increases the reliability in data transfer but adds one clock cycle latency in the system also. The coded data has been decoded in the following three steps: determination of the error locater polynomial, detection of error location using Chien Search Algorithm~\cite{chien:search} and location of the data at the error position. One can find the details of BCH algorithm in~\cite{chien:search}. In our present work we have designed the BCH encoding/decoding block as a custom hardware design. Instead of selecting BCH code with larger block size like (31, 26, 1) or (63, 57,1), we used eight BCH (15,7,2) in parallel for faster error correction without compromising the time complexity. Hence, each BCH encoder block can correct up to 2 bit of error within 7 bit of input. So the total  $8\times 2 = 16$  bit can be corrected using this technique with out any extra resources. Similarly, we can use triple error correcting BCH code~\cite{chien:search} but that will reduce the code rate.
\vspace{-1 pc}
\subsection{Interleaver/De-interleaver}To reduce the effect of burst error in the consecutive bytes of data in transmission process, we added interleaving block. Two types of interleaving strategies are there in any communication system viz block interleaver and convolutional interleaver. Here we have used block interleaver. Output of 120 bit data from encoder is divided into two block of 60 bit data and then interleaving operation is done on each 60 bit data using block interleaver. This block increases the code correction capabilities without any clock latency and overhead. De-interleaving process is used to reorder the data again in the receiver side.
\vspace{-1 pc}
\subsection{MUX/DEMUX and Clock Domain Crossing} 
\begin{figure}[h]
\begin{minipage}{14pc}
\includegraphics[width=22pc]{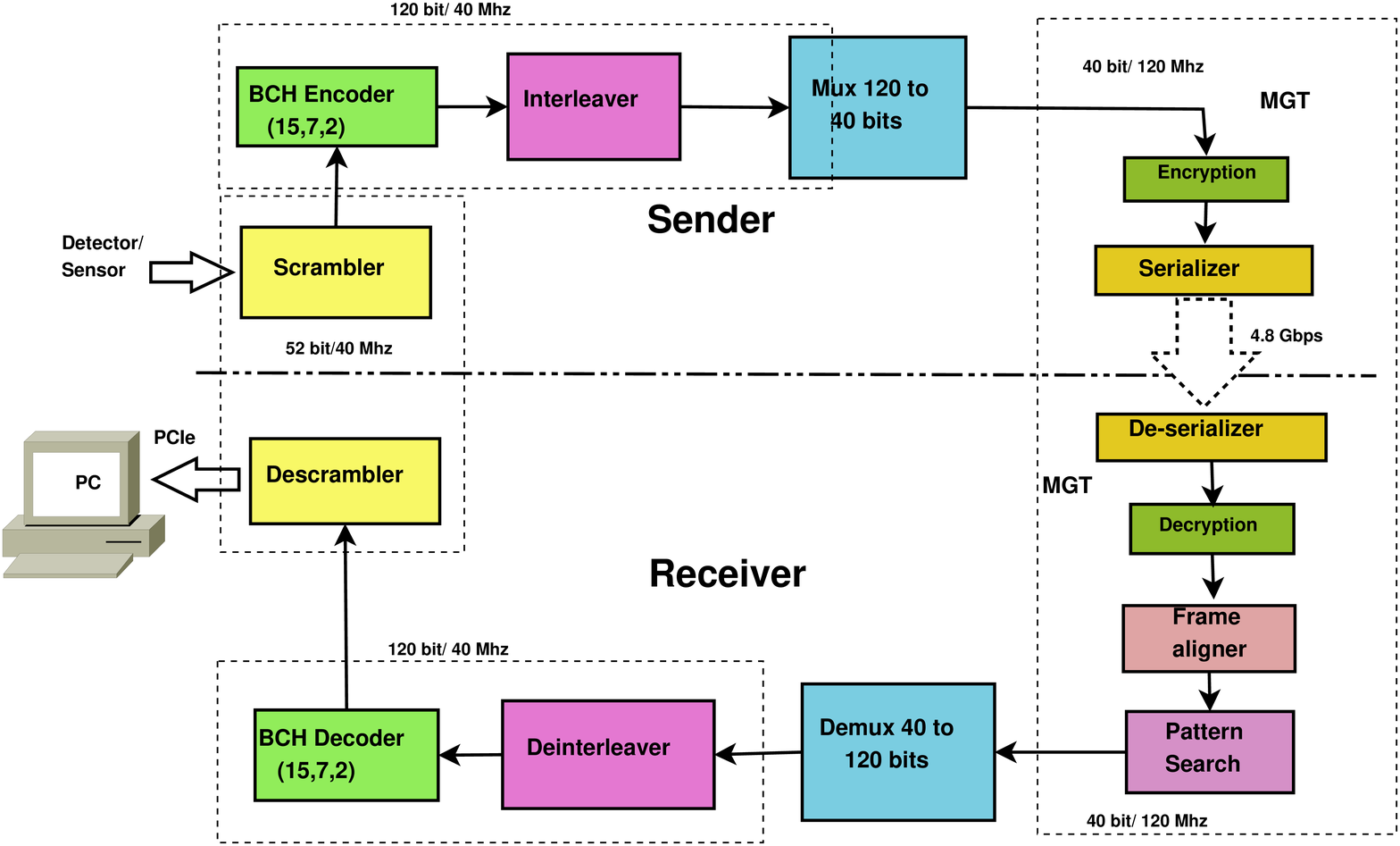}
\caption{\label{fig:BlockDiagramFlow}Internal blocks of the proposed system}
\end{minipage}\hspace{10pc}%
\begin{minipage}{14pc}

\includegraphics[width=20pc]{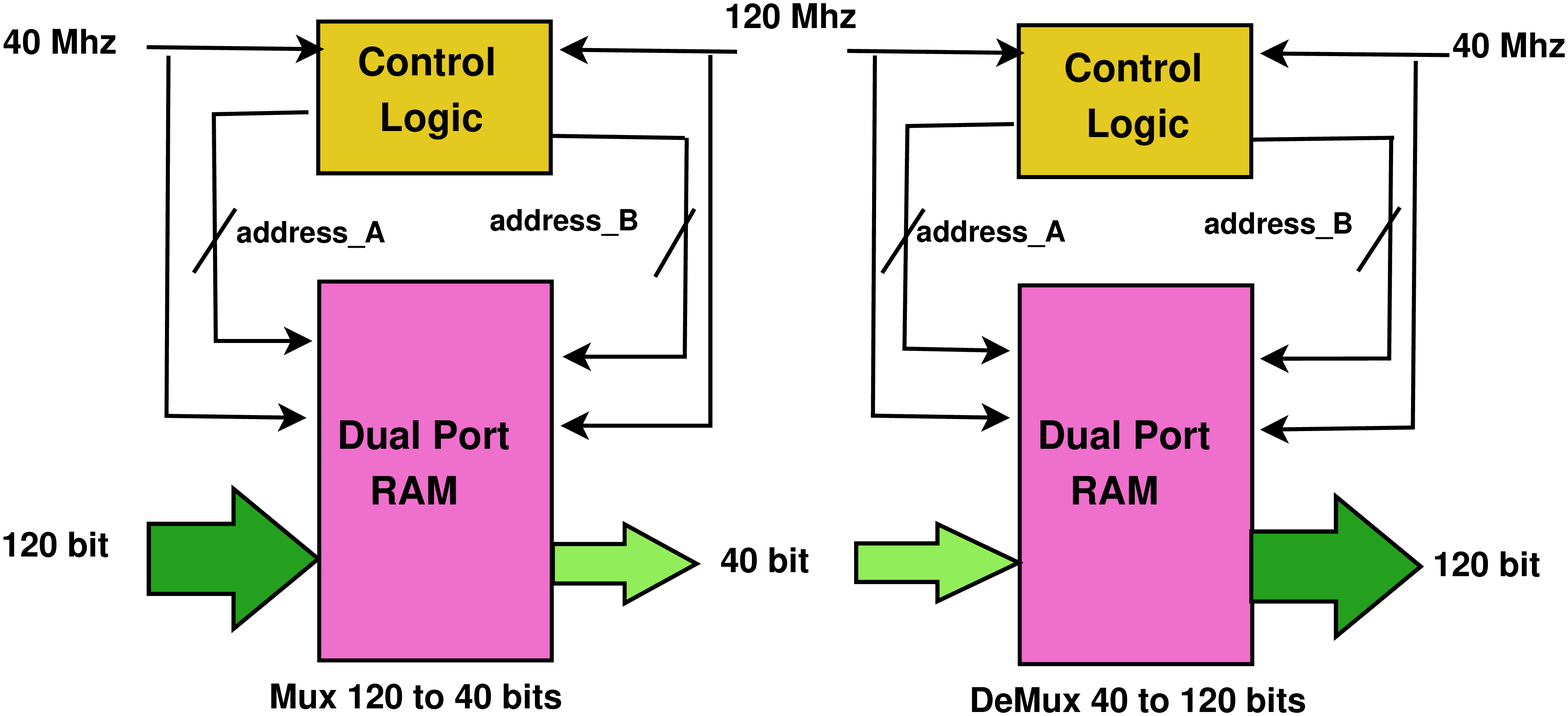}
\caption{\label{fig:MuxDMux}Mux DeMux for clock domain crossing}
\end{minipage} 
\end{figure}

This block consists of dual port RAM and read-write controller. It breaks down 120 bit frame into three words of 40 bit width. It reduces bandwidth consumption keeping the data rate same. Here, we have used 120 MHz clock to drive the multi-gigabit transceiver (MGT) available from Xilinx IP core to keep the data rate same with the internal blocks those are running with 40 MHz frequency. The data rate and clock frequency can be changed to any value according to the requirement. This block is used to synchronize the data rate between MGT and the other parts of the design. Figure~\ref{fig:MuxDMux} shows the architectural block diagram of the MUX-DEMUX and clock domain crossing.
%

\vspace{-1 pc}
\subsection{Encryption/Decryption} \label{security}
\begin{figure}[h]
\begin{minipage}{14pc}
\includegraphics[width=20pc]{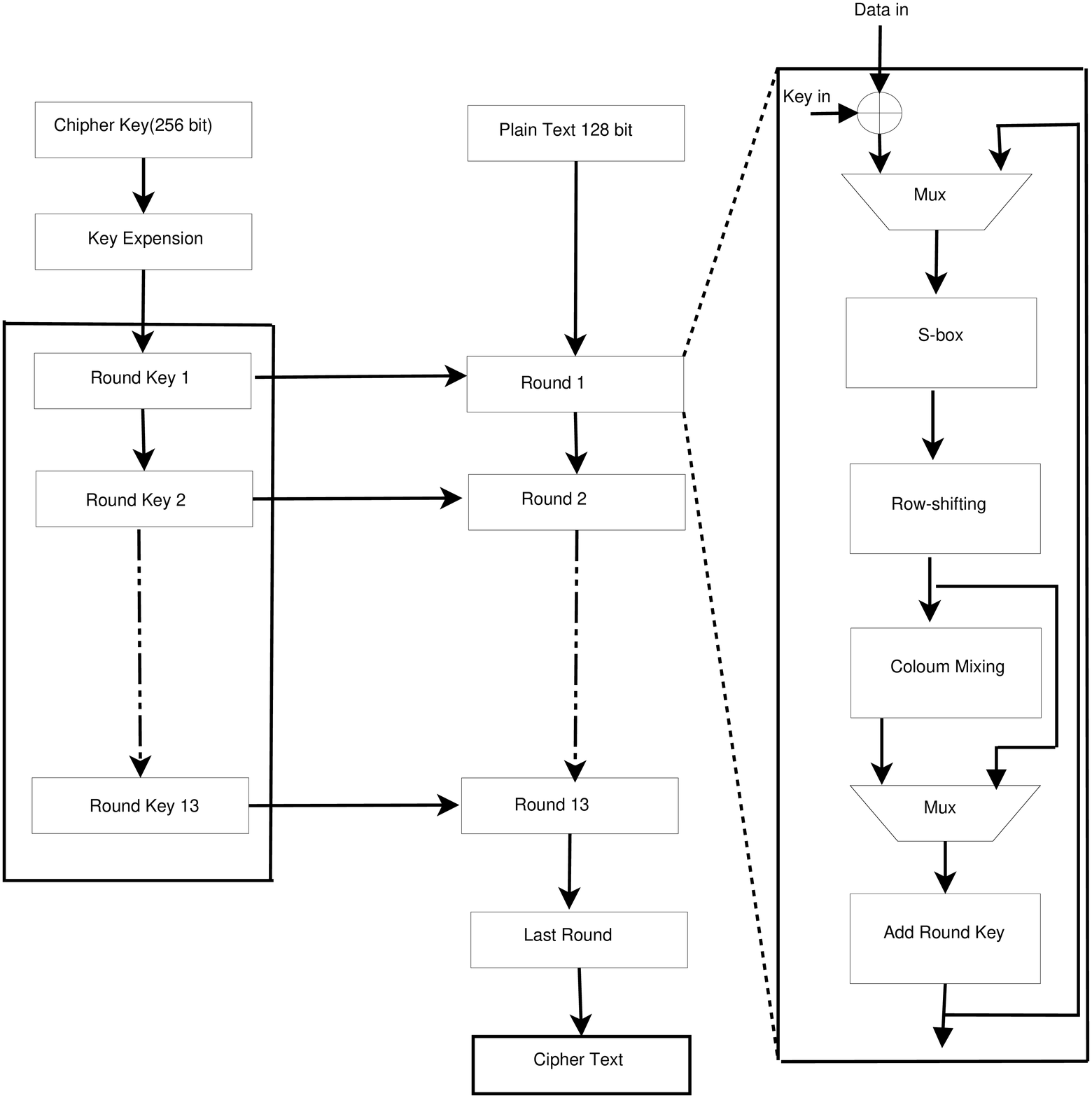}
\caption{\label{fig:aesencryption}Encryption flow of the AES algorithm}
\end{minipage}\hspace{7pc}%
\begin{minipage}{14pc}
\vspace{1pc}
\includegraphics[width=20pc]{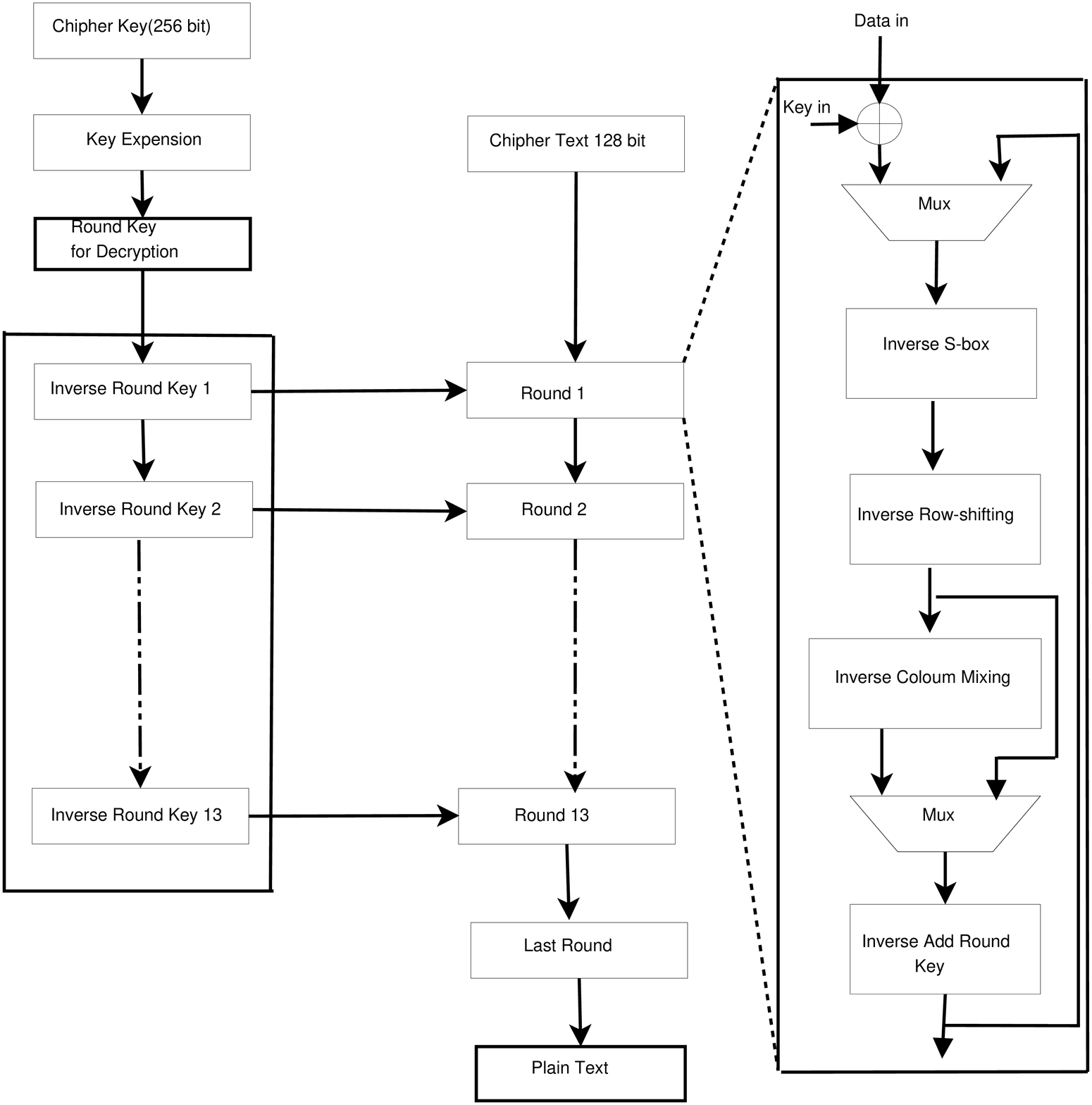}
\caption{\label{fig:aesdecryption}Decryption flow of the AES algorithm}
\end{minipage} 
\end{figure}

Encryption block is used to cipher the data, which will be transmitting over the channel. Different cryptographic algorithms may be used for this purpose. In our design, we tested the system with two different algorithms AES~\cite{aes:nist} and RSA~\cite{RSA:algo}. AES is a symmetric key cryptographic algorithm. The key size of AES algorithm can vary from 128,192 and 256 bits with fixed data size of 128 bits. Depending on the key size value of encryption and decryption operation, rounds may vary within in 10,12 and 14 respectively. Where as RSA works on asymmetric key based cryptographic algorithm. Two keys are used namely public and private key. Public key is used to encrypt the data whereas private key is used for decrypt the cipher data. There is no limitation on key size. Decryption receiver side is used for decipher the received data. Figure~\ref{fig:aesencryption} and Figure~\ref{fig:aesdecryption} shows the AES-256 encryption and decryption flow respectively.

\vspace{-1 pc}
\subsection{Serializer/De-serializer} To convert the parallel data to serial data, which is transmitted over the communication channel, serializer block is used. Similarly, De-serializer simply converts the serial data to parallel data in the receiver side.
\subsection{Frame Aligner and Pattern Search} \label{FrameAligner}
\begin{figure}[h]
\begin{minipage}{14pc}\vspace{-5pc}
\includegraphics[width=18pc]{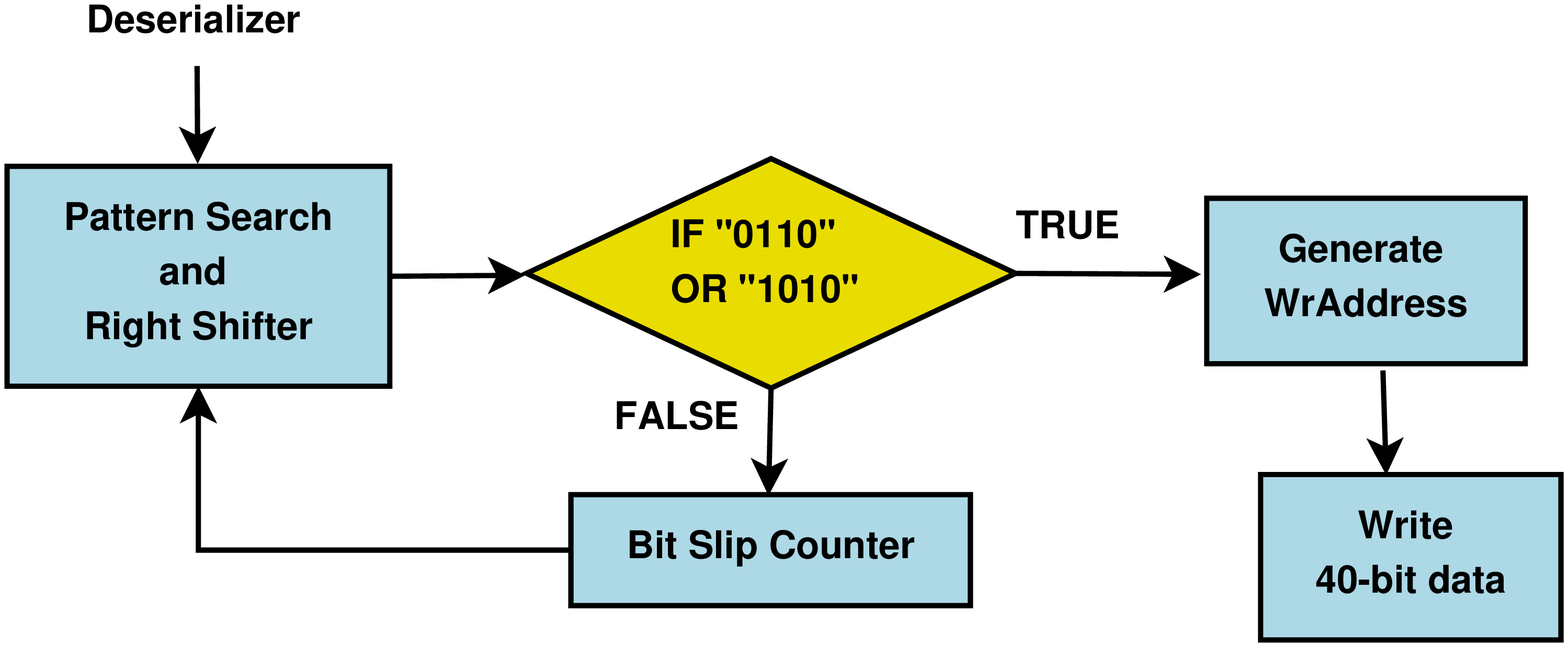}
\caption{\label{fig:FrameAlignerFlow}Algorithm for Frame Aligner and Pattern Search}
\end{minipage}\hspace{6pc}%
\begin{minipage}{14pc}\vspace{-4pc}
\vspace{3pc}
\includegraphics[width=18pc]{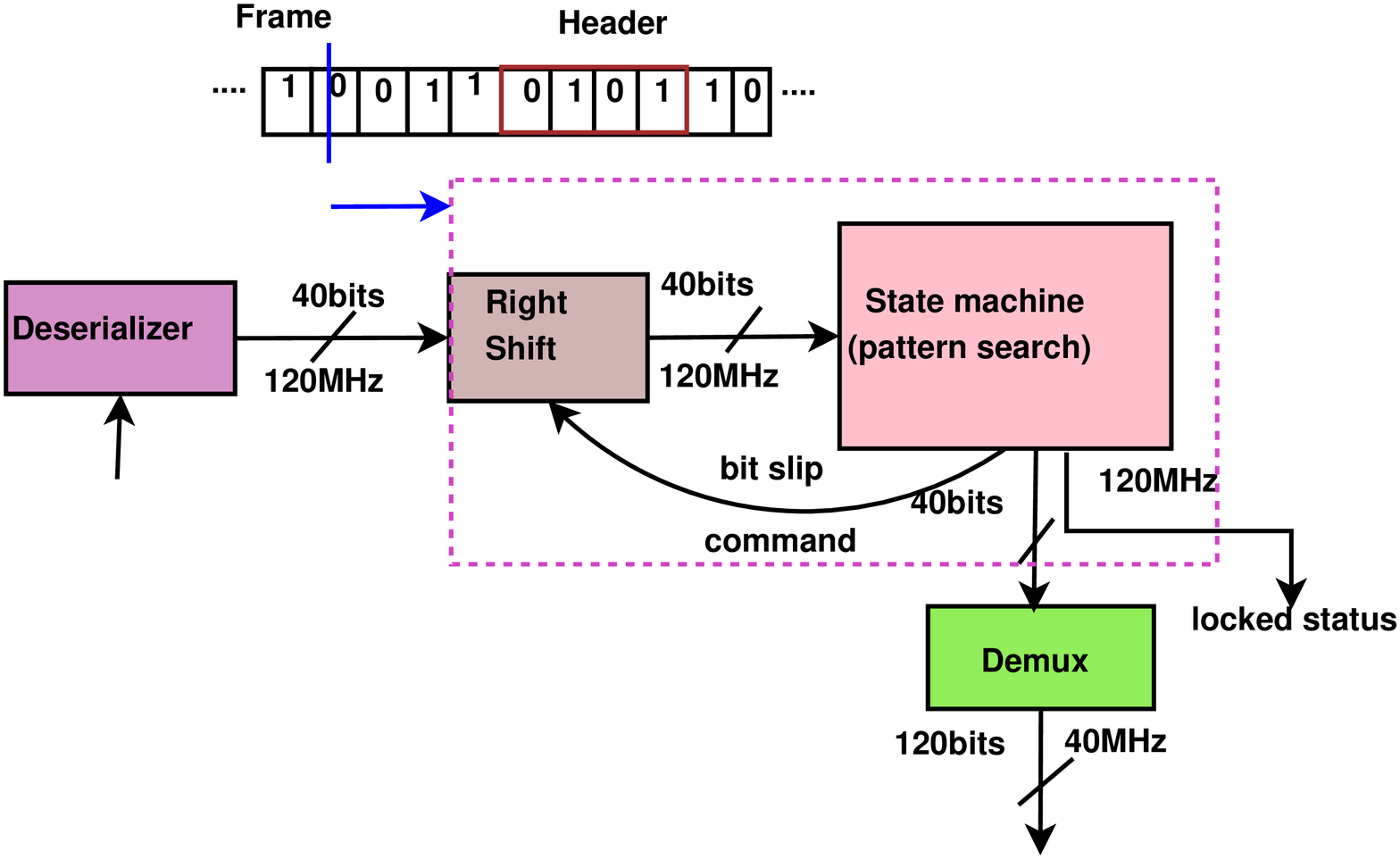}
\caption{\label{fig:FrameAlignerWork}Data flow diagrams of the Frame Aligner and Pattern Search block}
\end{minipage} 
\end{figure}

Frame aligner block is used in the receiver side for aligning and indexing the frames in a proper order. Pattern search algorithm is used to detect the frame header. Figure~\ref{fig:FrameAlignerFlow} shows the flow chart of this block. Two types of frames are considered in this design: standard frame and frame without FEC. 
Standard frame consists of four fields: Header field (4 bit), Slow Control (4 bit), Data field of width 48 bit, FEC field of width 64 bit. Whereas frame format without FEC consists all field of standard frame  except the FEC field. Header value of standard frame format and frame format without FEC are $1010$ and $0101$ respectively. The frame aligner and pattern search block consists of two sub blocks (Pattern search block, Right shift block) as shown in the Figure~\ref{fig:FrameAlignerWork}.

\vspace{-1 pc}
The pattern search block is used to check the whether the header field is properly received or not. It will be stable after a continuous search for another $32$ subsequent headers of other frames, after the first frame header is received.


\vspace{-1 pc}
\subsection{Data Transfer to Host PC through PCIe}
\begin{figure*}[htb]
\hspace{-2pc}
\includegraphics[scale=0.22]{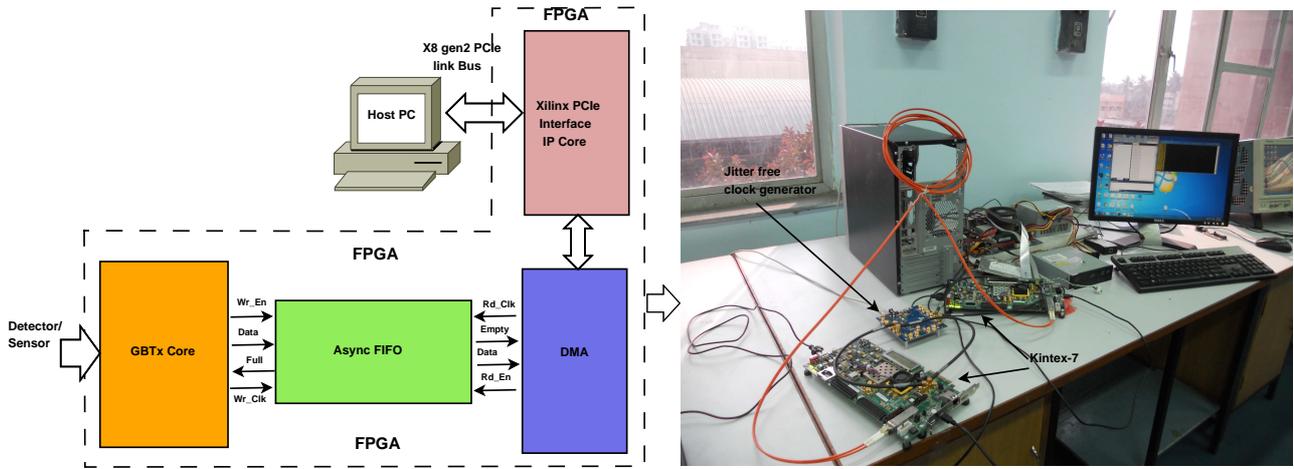}
\vspace{-5 pt}
\caption{ PCIe interfacing with blocks and Experimental setup of proposed DAQ}
\label{fig:PCIeSetup}

\end{figure*}

Transfer of data from FPGA board to PC is done by PCIe with the help of asynchronous Fast In Fast Out (FIFO) and SGDMA. We have used PCIe gen 2 IP core available from Xilinx. Interconnection of FPGA to PC through PCIe is shown in Figure~\ref{fig:PCIeSetup}. Data is written into FIFO at a frequency of 120 MHz and  data will be read from FIFO at a frequency of 125 MHz by which PCIe core is running. A program is written using windows software development kit (SDK) and C language for further processing of data in the PC.  
\vspace{-1 pc}
\subsection{Overview of Secure Data Flow }
\begin{figure*}[htb]
\hspace{-3 pc}
\includegraphics[scale=0.23]{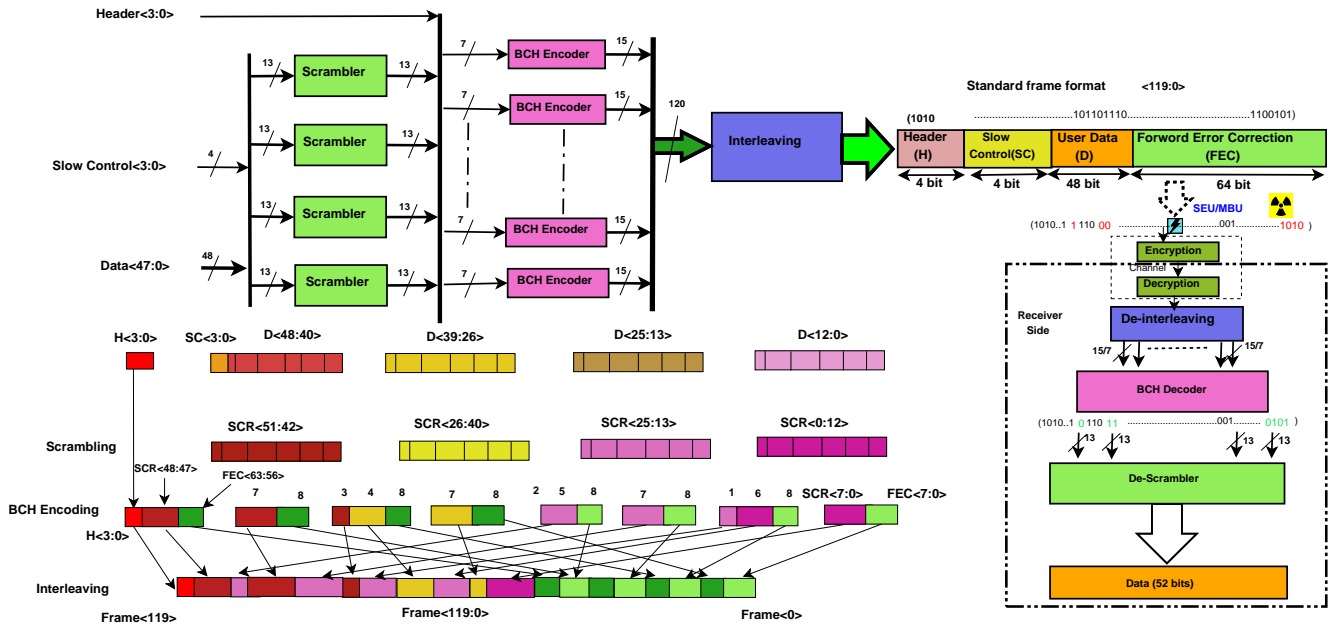}
\vspace{-10 pt}
\caption{Standard frame generation and Error correction flow with encryption/decryption}
\label{fig:data_flow}
\vspace{-15 pt}
\end{figure*}

 The complete chain of the functional blocks as shown in Figure~\ref{fig:BlockDiagramFlow} for the high speed secure DAQ with multi-bit error correction (Considering two bits error correction) has been implemented on the FPGA board. Figure~\ref{fig:data_flow} shows the complete mechanism of standard frame generation and the error correction flow with encryption/decryption mechanism.

52 bit of data received from detector is divided into four 13 bit block of data and scrambles each block parallely. The scrambled data with the 4 bit header is mapped in the input line of the eight BCH (15,7,2) encoders, which are running parallel.  Output of all the encoders are combined to get a frame of 120 bit data. This 120 bit of data are interleaved first and then goes to the next functional block that is the MUX. Interleaving is used to reduce the effect of burst error. But the header position is never changed in the frame format (red color in Figure~\ref{fig:data_flow}) even after interleaving process, helps to synchronize the frame in the receiver side. A encryption block is added after interleaving process for secure the data communication in the channel. Here, we use RSA and AES-128 algorithms for testing of our design. Details of the resource utilization of this block is described in section~\ref{PerformenceEvoluation}. In Mux-DeMux and clock domain crossing block, a dual port RAM is used to write 120 bit of data using 40 MHz clock and read the same data at 120 MHz clock rate with 40 bit word size. So the data rate for writing ($40 \times 120=4.8 $ Gbps) and for reading ($120\times 40=4.8$ Gbps) is same. The 40 bit data is serialized first and goes to the transmitter (TX) for transmitting over the optical fiber cable. In the receiver (RX) side functional blocks are Deserializer, DEMUX, Decryption, De-interleaver, BCH Decoder (15, 7, 2) and Descrambler. They perform reverse function with respect to Serializer, MUX, Encryption, Interleaver, BCH Encoder (15, 7, 2) and Scrambler respectively. The extra block frame aligner and pattern search in the receiver side is added in this chain whose functional description has been described in section~\ref{FrameAligner}.

\vspace{- 1 pc}
\section{Experimental Setup and performance Analysis} \label{PerformenceEvoluation}
The full prototype of the secure communication chain is implemented in the Xilinx Kintex-7 boards (KC705 from Avnet) using the Xilinx ISE 14.5 platform and VHDL for design entry. We have used an external jitter cleaned clock source (CDCE62005EVM of TI) to drive MGT of two Kintex-7 boards. One Agilent power supply was used to drive the whole system. Two Kintex-7 boards are connected through single mode optical fiber using Small Form-factor Pluggable (SFP) from \textit{Finisar} (FTLX8571D3BCL). For board to PC communication we have used eight lane PCIe gen2. 
The block diagram and experimental setup of the system are shown Figure~\ref{fig:PCIeSetup}. We achieved maximum bit rate of 4.8 Gbps in our system. In standard mode, a frame contains only 52 bit of data, 64 bit for error correction (FEC) and 4 bit of header. 64 bit for FEC can correct up to 16 bit of error, as it is applied on 8 encoder blocks in parallel (2 bit error correction for each block). So the data rate achieved considering only the data field (D) in this mode is: \\
$ 40 MHz \times 52 bit = 2.08 Gbps$ \\
In frame format without FEC, where error correction code is not used, the frame can carry $(52+64=116)$ bit of data out of 120 bit frame.
So in this mode data rate is measured: \\
$40 MHz \times 116 bit = 4.64 Gbps$ \\
Hence, the data transfer efficiency for the above mention two modes are $(2.08/4.80)\times 100 = 43.33\%$ and $(4.64/4.80 = 96.6\% $ respectively. Our system gives better performance in comparison with the system described in~\cite{kadric:socc:2012},~\cite{haoxu:nss:2007:XX},~\cite {mattihalli:cecnet:2012},~\cite{bohm:nss:mic:2012}.
Resource utilization for each functional block of the system is given in Table~\ref{table:resource}.

\begin{figure}[h]
\begin{minipage}{18pc}
\vspace{-5 pc}
\includegraphics[width=25pc]{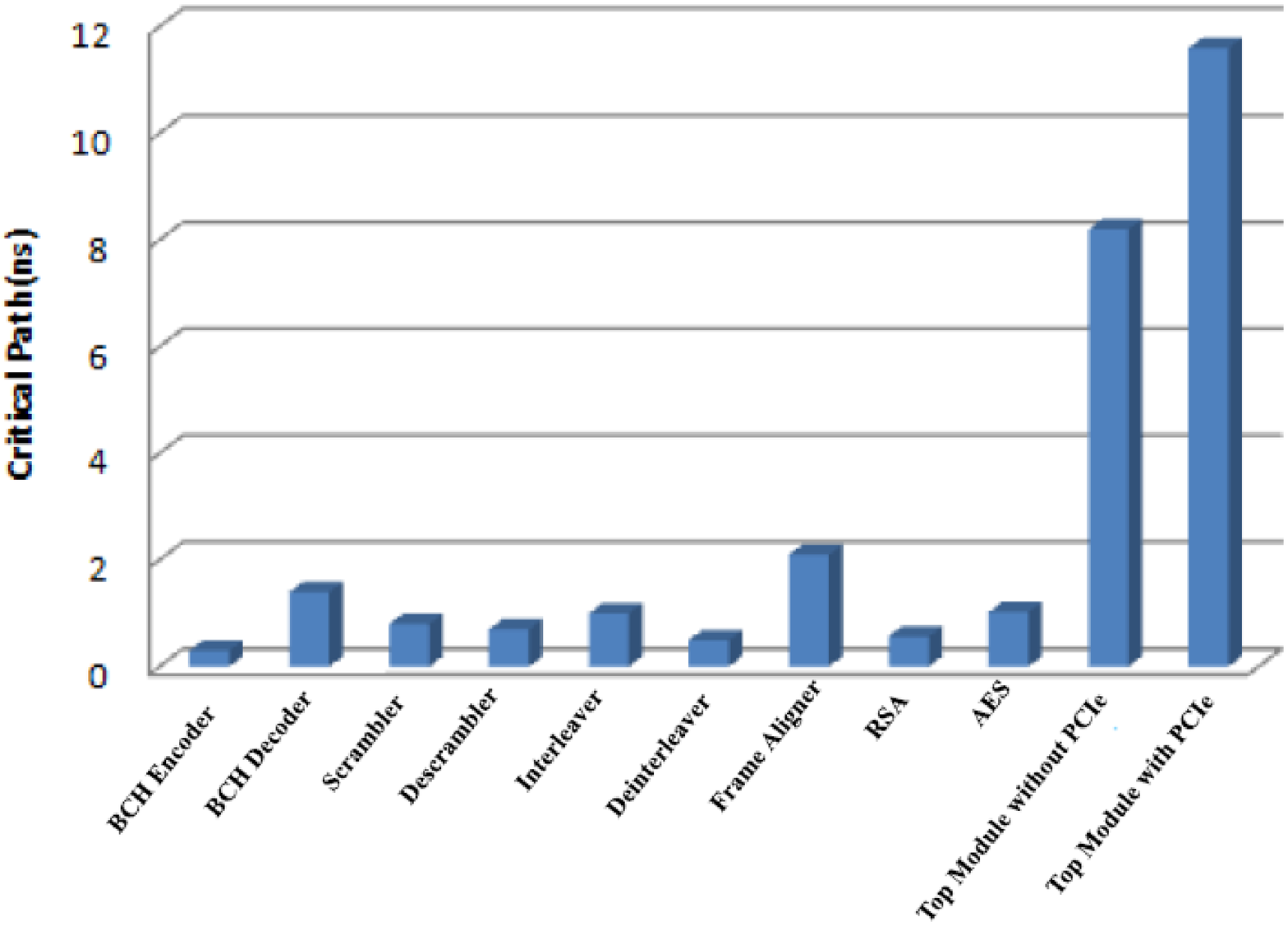}
\caption{\label{fig:CiticalPathDelay}Critical time for different block }
\end{minipage}\hspace{8pc}%
\begin{minipage}{14pc}
\vspace{-5 pc}
\includegraphics[width=18pc]{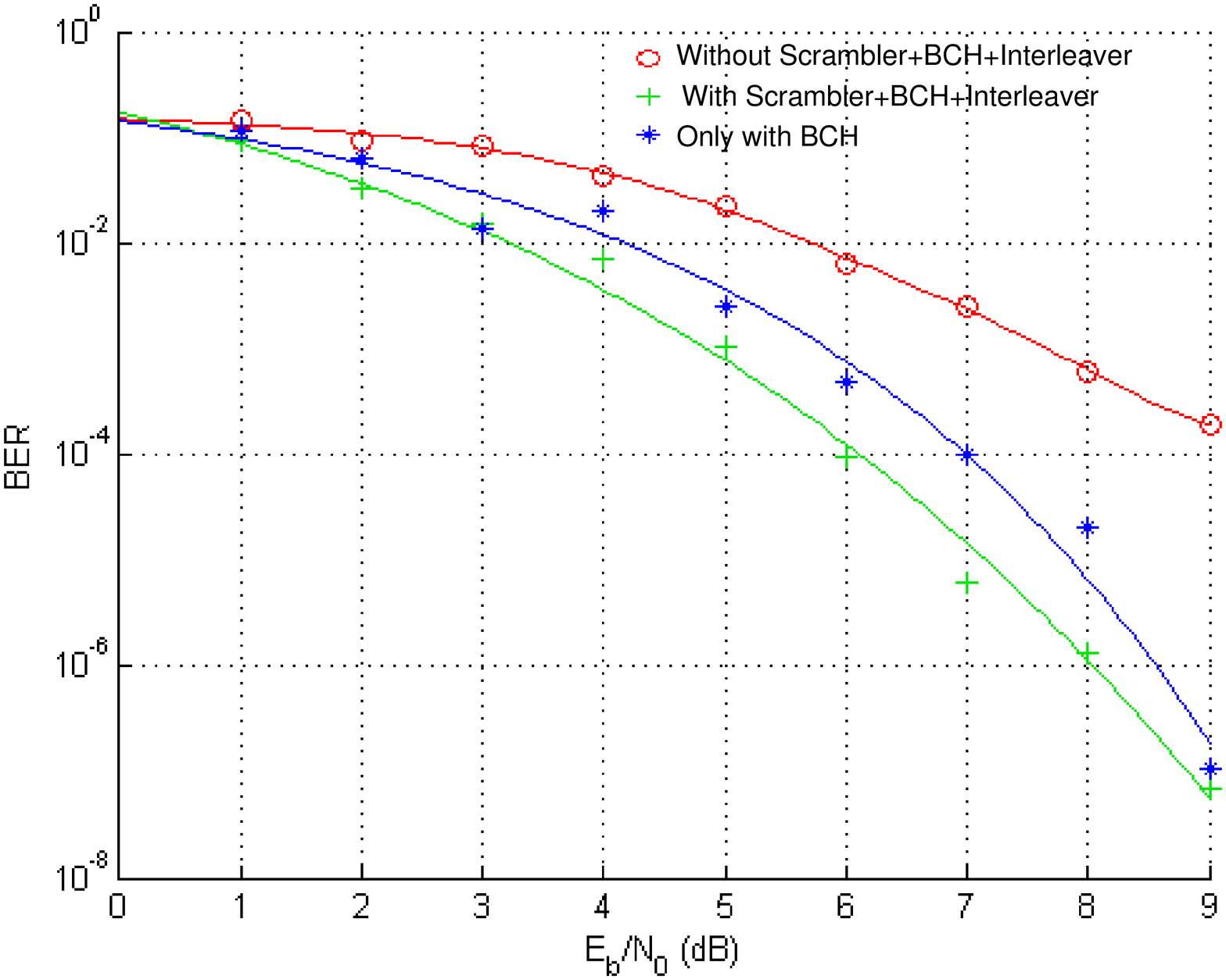}
\caption{\label{Rxmarginanalysis}Study of of BER with noise}
\end{minipage} 
\end{figure}

\begin{table}
\vspace{-3 pc}
\centering
\makebox[0pt][c]{\parbox{1.2\textwidth}{%
    \begin{minipage}[b]{0.56\hsize}\centering
       \caption{Resource Utilization}
\scalebox{0.6}{
\begin{tabular}{|c|c|c|c|c|c|}
\hline 
\textbf{Board} & \textbf{Module} & \textbf{Slice} & \textbf{Slice} &\textbf{ LUT} & \textbf{BRAM} \\ 
&\textbf{Name} &\textbf{Register} &\textbf{LUTs} & \textbf{Flip Flop}& \\
\hline 
\multirow{14}{*}{\rotatebox{90}{\textbf{Kintex 7- 325t}}} & BCH & 7//407600 & 951/203800& 0 &7/951 \\ 
& Encoder (15,7,2) &&&& \\
\cline{2-6}
 & BCH & 135/407600 & 446/203800 & 0 & 119/462 \\
& Decoder (15,7,2) &&&&(25\%) \\
 \cline{2-6}
 & Scrambler & 52 & 53 & 5 & 0 \\
 \cline{2-6}
  &Descrambler &104 & 56 & 5 & 0 \\
 \cline{2-6}
&Interleaver &44 & 40 & 40 & 0 \\
 \cline{2-6}
&DeInterleaver &201 & 82 & 80 & 0 \\
 \cline{2-6}
&Frame Aligner &115& 308 & 72 & 0 \\
 \cline{2-6}
 &Encryption (AES) &1311& 4300 & 864 & 0 \\
 \cline{2-6}
  &Encryption (RSA) &116	& 31612 & 75 & 0 \\
 \cline{2-6}
 &PCIe &5882 & 5287 & 2694 & 10 \\
 \cline{2-6}
 &Top Module&3665 &9003 & 1998 & 5 \\
  & Without PCIe & & &  &\\
 \cline{2-6}
 &Top Module&8360 & 8555 & 3779 & 26 \\
 & With PCIe & & & & \\
 \cline{2-6}  
 \hline
\end{tabular} }
\vspace{-6 pc}
\label{table:resource}
    \end{minipage}
        \hfill     
    \begin{minipage}[b]{0.42\hsize}
        \centering    
\caption{Module wise power consumption}
\scalebox{0.6}{

\begin{tabular}{|c|c|c|c|}
\hline 
\textbf{Board} & \textbf{Module } & \textbf{Logic } &\textbf{Signal}\\ 

&\textbf{Name} &\textbf{Power(mW)} & \textbf{Power(mW)}\\
\hline 
\multirow{16}{*}{\rotatebox{90}{\textbf{Kintex 7-325t}}} & BCH  & 0.02 & 0.01\\
& Encoder(15,7,2)&& \\
 \cline{2-4}
 & BCH  & 0.05 & 0.07 \\
&Decoder(15,7,2)&& \\
 \cline{2-4}
 & Scrambler & 0.04 & 0.00 \\
 \cline{2-4}
  &Descrambler &0.01 & 0.00 \\
 \cline{2-4}
&Interleaver &0.01 & 0.01\\
 \cline{2-4}
&DeInterleaver & 0.01& 0.02 \\
 \cline{2-4}
&Frame Aligner &1.34& 1.07 \\
 \cline{2-4}
&Encryption (RSA) &2.37& 1.57 \\
 \cline{2-4}
&Decryption (RSA) &3.64& 1.89 \\
 \cline{2-4}
&Encryption (AES) &2.76& 1.41 \\
 \cline{2-4}
&Decryption (AES) &2.90& 1.29 \\
 \cline{2-4}
 &PCIe &253.24 & 45.55 \\
 \cline{2-4}
 &Top Module& 474.18 & 2.91 \\
  & Without PCIe & &  \\
 \cline{2-4}
 &Top Module& 304.24 & 56.31 \\
 & With PCIe & &  \\
 \cline{2-4}  
 \hline
\end{tabular} }
\label{table:powerconsumtion}

    \end{minipage}
    \hfill    
}}
\end{table}
In Figure~\ref{fig:CiticalPathDelay} we show the critical time, which is the maximum delay time to get the output of a circuit for each of the circuit blocks. Power consumption is estimated using Xilinx Xpower tool and we show the estimated average logic and signal power for the various model of the proposed design. To the best of our knowledge, we are reporting the critical time and power consumption of this type of system for the first time.
\par SEU error is random in nature. We have emulated the SEU error by generating random error on the input data using random error generator~\cite{Antoni:IEEETran:2003}. The simulation results of BER is shown in the Figure~\ref{Rxmarginanalysis} with respect to the noise (Eb/N), which varies from 0 dB to 10 dB. Here we used Poisson distributed noise. Figure~\ref{Rxmarginanalysis} shows the efficiency of our system comprising of BCH code with interleaver,  scrambler, gives the best performance in presence the noise. The throughput of the DAQ system is measured as 4.8 Gbps using in Xilinx platform installed in Fedora OS.
%
%

%
%

\vspace{-1 pc}
\section{Conclusion}\label{Conclusion}
In this work we have proposed a DAQ design with fault tolerant secure communication for HEP experiments. The proposed DAQ supports high speed (in terms of Gbps) optical data communication and also corrects multi-bit error. In order to achieve secure communication, we have used AES-128 and RSA algorithms. The design has been implemented on Xilinx Kintex-7 board and real test setup has been developed involving board to board communication and PCIe interfacing with a host PC. A detailed performance analysis of the design implementation is presented in terms of timing diagram, resource utilization and critical timing for of each of the blocks (FPGA), power consumption and BER.
\vspace{-1 pc}
\ack
The authors like to acknowledge  DAE-DST (Govt of India), GSI, CERN, TEQIP-II (CU) to provide necessary support for carrying out the research.

\section*{References}
\bibliographystyle{unsrt}
\bibliography{iopart-num}


\end{document}